\documentclass[aps,prl,reprint,nobibnotes,nofootinbib]{revtex4-2}

\usepackage{amsmath,amssymb,mathtools}
\usepackage{hyperref}
\hypersetup{
    colorlinks=true,
    linkcolor=blue,
    citecolor=blue,
    urlcolor=blue}

\begin{document}

\title{Can a Local Quantum-Mechanical Description of Physical Reality Be Considered Complete?}

\author{E.~J.~Thompson}
\thanks{Corresponding author (Ethan Thompson)}
\email{thom3471@mylaurier.ca}
\affiliation{Wilfrid Laurier University, Waterloo, Canada, N2L 3C5}

\author{J. W. Moffat}
\affiliation{Perimeter Institute for Theoretical Physics, Waterloo, Ontario N2L 2Y5, Canada}
\affiliation{Department of Physics and Astronomy, University of Waterloo, Waterloo, Ontario N2L 3G1, Canada}

\date{\today}

\begin{abstract}
    In a complete quantum theory, one would construct it based on what we can observe in a system and operationally define in our universe. This philosophy goes back to some of the original ideals of quantum theory defined by W. Heisenberg, N. Bohr, and others as part of the Copenhagen school of thought. We argue in this letter that that if one wishes to construct a complete quantum field theory, that it is most natural to construct it based on a nonlocal theory of observable principles. If this interpretation is valid, then the one-particle nonrelativistic limit that recovers quantum mechanics would as well be dynamically nonlocal\footnote{Note that Bell nonlocality describes nonlocal correlations between distant entangled particles that do not alter local probabilities or allow communication, while dynamical nonlocality involves nonlocal influences on a single particle's modular equations of motion for example the Aharonov-Bohm effect}. In this interpretation we derive the nonlocal Schr\"odinger equation. We will as well demonstrate that the canonical position--momentum commutator and the Heisenberg uncertainty principle emerge from translation covariance.
\end{abstract}

\maketitle

In 1935 A. Einstein, B. Podolsky, and N. Rosen asked if the wavefunction could be considered a complete physical description of reality. To them this meant that the wavefunction should contain a complete specification of those elements of reality that may be assigned to physical observables of a system~\cite{EPR1935}. N. Bohr replied to EPR later that same year, arguing that their criterion of physical reality contains an essential ambiguity when applied to quantum phenomena and that quantum mechanics, understood through complementarity and the experimental conditions defining an observable phenomenon, satisfies the appropriate demands of completeness~\cite{Bohr1935}. His response reflected a broader methodological principle of the Copenhagen approach, that physical meaning should be attached to quantities through the conditions under which they can be operationally defined, rather than by assuming that every mathematical quantity corresponds to a context-independent property of reality~\cite{Bohr1928}. Bohr famously stated\footnote{We do not know when this was stated, it was attributed to him by his close associate and student, Aage Petersen, who published the quote in a 1963 article titled \emph{The Philosophy of Niels Bohr}.}: ``It is wrong to think that the task of physics is to find out how nature is. Physics concerns what we can say about nature.'' Bohr was arguing that science relies on what we can measure, test, and observe in experiments. In his view, words and math are human tools to map experiences, not a mirror of an ultimate reality. Bohr argued there is only an abstract physical description, not a literal subatomic realm we can picture.

Einstein deeply disliked the Copenhagen interpretation because it abandoned two principles he considered non-negotiable for physics, determinism and realism. 

Realizing he couldn't prove quantum mechanics wrong, Einstein tried to prove it incomplete, this was done along with Podolsky and Rosen, he showed that quantum mechanics allowed two particles to become "entangled." Measuring one particle instantly determined the state of the other, even if they were light-years apart. Einstein called this "spooky action at a distance" and argued that the particles must have carried hidden, predetermined traits all along.

As an almost one-hundred year follow-up to this debate, we want to ask a related question:
\\
\\
\emph{Even if quantum mechanics is complete as a probabilistic theory, can its assumption of arbitrary locality be regarded as fundamental?}
\\
\\
The development of quantum theory repeatedly forced physics away from the classical description of particles as objects possessing simultaneously definite properties at mathematical points. We accept Bohr's operational challenge and apply it to locality itself.

Despite this, ordinary nonrelativistic quantum mechanics nevertheless keeps an idealization inherited from classical geometry. This position is formally  represented by arbitrarily sharp resolution. Position eigenstates and sharply bounded projection operators provide a mathematically useful description in which localization can, in principle, be refined without an intrinsic spatial scale.

The status of this assumption was already questioned explicitly by Dirac. In his 1948 analysis of ``localizable dynamical systems,'' Dirac defined a  relativistic quantum system to be localizable when its wave function could be represented entirely in terms of dynamical variables associated with physical conditions at individual spacetime points~\cite{Dirac1948}. Significantly, he did not assume that atomic systems must possess this property. Rather, he asked whether atomic systems are in fact localizable at all, and allowed that the atomic world might contain features that cannot be described in terms of localized dynamical variables. He further noted contemporary attempts to formulate theories in terms of quantities that were not accurately localizable~\cite{Dirac1948}. Dirac did not conclude that such a nonlocalizable theory was required, but he  emphasized the considerable difficulty of constructing one. His analysis nevertheless isolated locality as an assumption about the physical 
description of a quantum system rather than an automatic consequence of 
quantum mechanics.

Relativistic quantum field theory already makes this assumption considerably less straightforward, since quantum fields are operator-valued distributions, physical observables are defined by suitable smearings~\cite{GardingWightman1964}, and positive-energy relativistic one-particle states possess well-known limitations on sharp localization~\cite{NewtonWigner1949,Hegerfeldt1974,Busch1999}, this is also done retroactivity and not proactively. One must therefore distinguish a coordinate point on the spacetime manifold from a physically realizable localization measurement. This idea is particularly sharp when one takes into account Hegerfeldt's result. He showed that for a Hamiltonian bounded from below, a state that is strictly localized within a bounded spatial region cannot, under general assumptions, remain compactly localized under relativistic time evolution. Instead, an initially localized state generically develops nonvanishing spatial tails immediately~\cite{Hegerfeldt1974}. This does not itself imply operational superluminal signalling, but it shows us that exact compact particle localization cannot straightforwardly be identified with a fundamental relativistic observable.

There is also a structural distinction between relativistic quantum field theory and ordinary nonrelativistic quantum mechanics that is important for the present argument. In QFT, fields are defined over spacetime and locality is formulated through relations between observables associated with spacetime regions. In nonrelativistic quantum mechanics, time enters as an external evolution parameter, while position is represented within the Hilbert-space observable structure. The notion of localization appearing in quantum mechanics is not an independent spacetime structure, but is inherited from the one-particle reduction of the underlying field theory. If the physical spacetime localization structure of that parent theory is fundamentally nonlocal, its one-particle quantum-mechanical limit should inherit that nonlocality in its physical position observables.

This distinction becomes fundamental in a UV complete nonlocal quantum field theory. In the class considered here, fields and observables are regulated by an entire function of the covariant d'Alembertian~\cite{Moffat1990,Evens1991,Moffat2019,MoffatThompsonRegulators}, with a corresponding nonlocal energy scale, $E_M=\hbar c(L_M^{-1})$, called the Moffat energy-length. For an admissible entire-function regulator the theory preserves finite physical pole structure, while modifying UV propagation and interactions~\cite{MoffatThompsonRegulators}. In position space the same operator can be represented by a noncompact smearing kernel associated with the Moffat length. The physical observable associated with a spacetime point is thus not point supported, but nonlocal over a region of size $L_M$~\cite{ThompsonLocalization,ThompsonPhaseSpace}.

For clarity, we distinguish this notion of nonlocality from Bell nonlocality. Bell nonlocality refers to correlations between measurements on entangled systems that cannot be reproduced by a local hidden-variable theory, and such correlations can occur even in an otherwise local relativistic quantum field theory~\cite{Bell1964}. Such quantum correlations and their relation to nonlocal quantum field theory and entanglement are investigated in ref.~\cite{LandryMoffat2024}. The nonlocality considered here is instead a modification of the physical localization and interaction structure of the theory itself, where observables and interactions are not fundamentally point supported, but possess an intrinsic spacetime extent characterized by $L_M$. The Bell nonlocality concerns the structure of quantum correlations, whereas the present nonlocality concerns the spacetime support and operational localization of physical observables.

Over the past decades there has been a substantial amount of work developing nonlocal quantum theories. One is led to the simple yet nuanced question: if ordinary quantum mechanics arises as the nonrelativistic one-particle sector of quantum field theory, what quantum mechanics follows when the underlying field theory is itself fundamentally nonlocal? 

We will derive the one-particle limit of nonlocal quantum field theory, derive the nonlocal Schr\"odinger equation. We will demonstrate that the canonical position--momentum commutator and the Heisenberg uncertainty principle emerge from translation covariance rather than being imposed independently. In this sense, if the underlying quantum field theory is nonlocal, exact local quantum mechanics is not discarded, but rather it is recovered as the infrared approximation to a more general quantum description.

\emph{The one particle limit of nonlocal quantum field theory} -- We now establish the relation between the underlying nonlocal quantum field theory and its one-particle quantum-mechanical limit. For clarity we consider a massive real scalar field \footnote{The essential argument depends only on the existence of a stable one-particle pole and generalizes to fields with spin.}. The free nonlocal equation of motion is given by:
\begin{equation}
    \mathcal{F}^{-1}\!\left(\frac{\Box}{E_M^2}\right)
    (\Box+m^2)\phi=0.
\end{equation}
Because $\mathcal F$ has no finite zeros, the regulating operator is invertible on the physical domain.

The entire-function deformation does not introduce a new free dispersion relation or new finite-energy one-particle states. It changes off-shell propagation and ultraviolet weighting, while the physical one-particle pole remains that of the local theory~\cite{MoffatThompsonRegulators}.

It is important to distinguish this statement from a claim that all physical momentum observables remain local. The four-momentum $p^\mu$ labeling the plane-wave solutions is the eigenvalue of the global translation generator associated with Poincar\'e symmetry. Preservation of this generator is required by translation invariance and is compatible with a nonlocal observable algebra~\cite{Thompson2026Covariance}. A finite-resolution measurement of the momentum content of a nonlocal excitation need not coincide with the sharp spectral measure of this global generator. We will distinguish between the canonical momentum $P$, which generates translations, and the physical momentum observable $P_M$, which may inherit the nonlocal response of the regulated field theory~\cite{ThompsonLocalization,ThompsonPhaseSpace}.

We can now derive the corresponding one-particle quantum mechanics. Expanding the positive-frequency part of the field gives us:
\begin{equation}
    \phi^{(+)}(x)
    =
    \int
    \frac{d^3p}
    {(2\pi)^{3/2}\sqrt{2E_{\mathbf p}}}
    a(\mathbf p)e^{-ip\cdot x},
\end{equation}
with $E_{\mathbf p}=\sqrt{m^2c^4+c^2\mathbf p^2}$. We then consider the normalized one-particle state:
\begin{equation}
    |\Psi\rangle
    =
    \int d^3p\,
    \widetilde{\psi}(\mathbf p)
    a^\dagger(\mathbf p)|0\rangle ,
\end{equation}
where $\widetilde{\psi}(\mathbf p)$ is the amplitude in the eigenbasis of the global translation generator. After absorbing the conventional relativistic normalization factors into $\widetilde{\psi}$, we find the associated one-particle amplitude. Then taking its time derivative gives us the exact positive-energy free one-particle evolution equation. 

To obtain the nonrelativistic sector, we separate the rapid rest-energy phase, and expand for $|\mathbf p|\ll mc$, then removing the rest energy gives us the Schr\"odinger equation with its first relativistic correction for a free particle:
\begin{equation}
    i\hbar\frac{\partial\chi}{\partial t}
    =
    -\frac{\hbar^2}{2m}\nabla^2\chi
    -
    \frac{\hbar^4}{8m^3c^2}\nabla^4\chi
    +\cdots .
\end{equation}

The ordinary free Schr\"odinger mechanics emerges from the stable, positive-energy one-particle sector of the relativistic field theory. The same conclusion holds for the free sector of the entire-function theory, because invertible nonlocal dressing leaves its mass shell and global translation generators unchanged~\cite{MoffatThompsonRegulators,Thompson2026Covariance}. Nonlocal quantum mechanics appears through the interactions and finite-resolution observables inherited from the nonlocal field algebra.

\emph{The Nonlocal Schr\"odinger Equation} -- Nonlocality must survive the one-particle reduction through interactions and physical observables rather than through an arbitrary modification of the free dispersion relation. This distinction will be important below as the canonical momentum $P$ remains the generator of translations, while a finite-resolution physical momentum observable $P_M$ may inherit the nonlocal response of the regulated field algebra. We first derive the corresponding nonlocal Schr\"odinger equation.

We consider a scalar field interacting with a static external potential~\cite{Frolov2021}. It is convenient to write the relativistic quadratic theory in the form:
\begin{equation}
    S
    =
    \int d^4x\,
    \left[
        \Phi^\dagger
        \mathcal{F}^{-1}\!\left(\frac{\Box}{E_M^2}\right)
        \mathcal{K}_0
        \Phi
        -
        \Phi^\dagger U\,\Phi
    \right],
\end{equation}
where $\mathcal{K}_0=\Box+m^2$, and $U(\mathbf x)$ is the relativistic interaction operator. For the nonrelativistic scalar potential $V(\mathbf x)$ we take, in units $\hbar=c=1$, $U(\mathbf x)=2mV(\mathbf x)$. The precise distribution of entire-function factors between kinetic terms, propagators, and vertices is convention dependent~\cite{Moffat1990,Evens1991,MoffatThompsonRegulators}. We introduce the square-root smearing operator:
\begin{equation}
    \mathcal{E}_M
    :=
    \mathcal{F}^{1/2}\!\left(\frac{\Box}{E_M^2}\right),
\end{equation}
which exists on the physical domain because $\mathcal F$ is entire and zero-free~\cite{MoffatThompsonRegulators}. We now perform the invertible field redefinition $\Phi=\mathcal{E}_M\Psi$. Since $\mathcal E_M$ is a function of $\Box$, it commutes with the free kinetic operator and thus $[\mathcal{E}_M,\mathcal{K}_0]=0$. The kinetic term therefore becomes:
\begin{align}
    \Phi^\dagger
    \mathcal{F}^{-1}\mathcal{K}_0
    \Phi
    &=
    \Psi^\dagger
    \mathcal{E}_M
    \mathcal{F}^{-1}
    \mathcal{K}_0
    \mathcal{E}_M
    \Psi
    \nonumber\\
    &=
    \Psi^\dagger
    \mathcal{K}_0
    \Psi,
\end{align}
because $\mathcal{E}_M \mathcal{F}^{-1} \mathcal{E}_M=1$. The interaction term, by contrast, becomes $\Phi^\dagger U\Phi=\Psi^\dagger\mathcal{E}_M U\mathcal{E}_M\Psi$.
So the nonlocality can be shifted from the free quadratic operator into the interaction~\cite{Evens1991,MoffatThompsonRegulators}. The relativistic one-particle equation is thus:
\begin{equation}
    (\Box+m^2)\Psi
    +
    \mathcal{E}_M U\mathcal{E}_M\Psi
    =
    0.
\end{equation}

For a static background we now pass to the equal-time spatial operator. For the Gaussian representative of the entire-function class we write~\cite{MoffatThompsonRegulators}:
\begin{equation}
    \mathcal{E}_M
    =
    \exp\!\left(
        \frac{L_M^2}{2}\nabla^2
    \right),
    \qquad
    L_M=\frac{\hbar c}{E_M},
\end{equation}
so that in the eigenbasis of the global spatial translation generator is:
\begin{equation}
    \mathcal{E}_M(\mathbf p)
    =
    \exp\!\left(
        -\frac{L_M^2\mathbf p^2}{2\hbar^2}
    \right)
    =
    \exp\!\left(
        -\frac{\mathbf p^2c^2}{2E_M^2}
    \right).
\end{equation}
where $\mathbf p$ is the eigenvalue of the canonical momentum operator $P$. To obtain the nonrelativistic limit, we take $\Psi(\mathbf x,t)=e^{-imt}\psi(\mathbf x,t)$ temporarily in natural units. Since $\partial_t^2\Psi=e^{-imt}\left(-m^2\psi-2im\partial_t\psi+\partial_t^2\psi\right)$,
the relativistic equation gives us $\partial_t^2\psi-2im\partial_t\psi-\nabla^2\psi+2m\mathcal{E}_MV\mathcal{E}_M\psi=0$. For a slowly varying nonrelativistic envelope, $|\partial_t^2\psi|\ll m|\partial_t\psi|$, and the first term may be neglected. Dividing the remaining equation by $2m$ gives us:
\begin{equation}
    i\partial_t\psi
    =
    -\frac{\nabla^2}{2m}\psi
    +
    \mathcal{E}_MV\mathcal{E}_M\psi.
\end{equation}
Restoring $\hbar$, we find:
\begin{equation}
    i\hbar\frac{\partial\psi}{\partial t}
    =
    \left[
        -\frac{\hbar^2}{2m}\nabla^2
        +
        \mathcal{E}_MV\mathcal{E}_M
    \right]\psi
.
\end{equation}
This is the nonlocal Schr\"odinger equation induced by the one-particle sector of the underlying nonlocal field theory. The kinetic operator retains the canonical translation generator, $-\hbar^2\nabla^2=P^2$, while the interaction is nonlocal.

The nonlocal character becomes explicit in coordinate space. Here the Gaussian operator possesses the heat-kernel representation:
\begin{equation}
    (\mathcal{E}_Mf)(\mathbf x)
    =
    \int d^3y\,
    K_M(\mathbf x-\mathbf y)f(\mathbf y),
\end{equation}
where:
\begin{equation}
    K_M(\mathbf r)
    =
    \frac{1}{(2\pi L_M^2)^{3/2}}
    \exp\!\left(
        -\frac{\mathbf r^2}{2L_M^2}
    \right),
\end{equation}
and:
\begin{equation}
    \int d^3r\,K_M(\mathbf r)=1.
\end{equation}
The interaction therefore acts as:
\begin{equation}
    (\mathcal{E}_MV\mathcal{E}_M\psi)(\mathbf x)
    =
    \int d^3y\,
    V_M(\mathbf x,\mathbf y)
    \psi(\mathbf y),
\end{equation}
with the nonlocal potential kernel:
\begin{equation}
    V_M(\mathbf x,\mathbf y)
    =
    \int d^3z\,
    K_M(\mathbf x-\mathbf z)
    V(\mathbf z)
    K_M(\mathbf z-\mathbf y).
\end{equation}
The wave equation takes the explicitly integral form:
\begin{equation}
    i\hbar\partial_t\psi(\mathbf x,t)
    =
    -\frac{\hbar^2}{2m}\nabla^2\psi(\mathbf x,t)
    +
    \int d^3y\,
    V_M(\mathbf x,\mathbf y)
    \psi(\mathbf y,t).
\end{equation}

Here, unlike the local interaction $V(\mathbf x)\psi(\mathbf x)$, the evolution at $\mathbf x$ now depends on the wave amplitude throughout a neighborhood of characteristic radius $L_M$. The one-particle excitation is therefore dynamically quasi-local rather than point-coupled~\cite{ThompsonLocalization}.

The same result is particularly transparent in momentum space. Eq.~\eqref{ME1},\eqref{ME2} are matrix elements of a nonlocal interaction potential. The general regulated interaction is given by:
\begin{equation}
    \langle\mathbf p|V_M|\mathbf p'\rangle
    =
    \mathcal{E}_M(\mathbf p)
    \widetilde V(\mathbf p-\mathbf p')
    \mathcal{E}_M(\mathbf p').
    \label{ME1}
\end{equation}
For the Gaussian specific form, we have:
\begin{equation}
    \langle\mathbf p|V_M|\mathbf p'\rangle
    =
    e^{-\mathbf p^2c^2/(2E_M^2)}
    \widetilde V(\mathbf p-\mathbf p')
    e^{-\mathbf p'^2c^2/(2E_M^2)}.
    \label{ME2}
\end{equation}
Large canonical-momentum components of interaction matrix elements are thus suppressed by the same nonlocal mechanism that softens ultraviolet behavior in the parent field theory~\cite{MoffatThompsonRegulators}. This suppression should not be confused with a modification of the translation generator itself. Rather, it describes how states of definite canonical momentum couple through a nonlocal interaction.

The ordinary local Schr\"odinger equation is recovered in the appropriate limit. Thus local quantum mechanics is not replaced, but obtained as the $L_M\rightarrow0$ limit of the nonlocal one-particle dynamics.

\emph{The Emergence of the Uncertainty Principle} -- We now derive the position--momentum uncertainty principle from the one-particle structure inherited from the underlying field theory.

For simplicity consider one spatial dimension. Since the entire-function deformation preserves continuous translations~\cite{Thompson2026Covariance}, the one-particle Hilbert space carries a strongly continuous unitary representation $U(a)$. By Stone's theorem~\cite{Stone1932}, there exists a self-adjoint generator $P$ such that $U(a)=\exp\left((-i/\hbar)aP\right)$. Translations act on the one-particle amplitude as $(U(a)\psi)(x)=\psi(x-a)$. Expanding both expressions to first order in $a$ and comparing terms gives us:
\begin{equation}
P=-i\hbar\frac{\partial}{\partial x}.
\end{equation}
Momentum therefore enters as the generator of spatial translations rather than as an independently postulated operator. Physical localization in the nonlocal theory need not be sharp. Let $E_X(dx)$ be a centered translation-covariant localization POVM~\cite{Busch1999}. Its first moment:
\begin{equation}
X_M=\int x,E_X(dx),
\end{equation}
coincides with the canonical position operator $X$ for a normalized centered response kernel, although its higher moments need not. Translation covariance therefore gives us: $U(a)^\dagger X U(a)=X+aI$. Using the unitary representation above and differentiating with respect to $a$ at $a=0$ gives us $[X,P]=i\hbar I$. So the canonical position--momentum commutator follows directly from the translation structure of the one-particle theory~\cite{Stone1932}.

The uncertainty relation then follows from positivity. For a normalized state we define $A=X-\langle X\rangle,\qquad B=P-\langle P\rangle$.
For every real $\lambda$ we have that $0
\leq
|(A+i\lambda B)|\psi\rangle|^2 \nonumber\
=
(\Delta X)^2+\lambda^2(\Delta P)^2-\lambda\hbar$. Since this quadratic must remain nonnegative for every $\lambda$, its discriminant cannot be positive. Thus:
\begin{equation}
\Delta X\Delta P\geq\frac{\hbar}{2}.
\end{equation}

The Heisenberg relation is thus not an additional assumption of the one-particle theory, but it follows from continuous translations, covariant localization, and Hilbert-space positivity. Nonlocality itself does not create quantum uncertainty. Its additional physical effect appears in the higher moments of the finite-resolution observables, which can acquire irreducible nonlocal widths even though their first moments and the canonical commutator remain unchanged.

\emph{Beyond Heisenberg: Nonlocal Physical Uncertainty} -- The preceding section established the ordinary position--momentum uncertainty relation~\cite{Heisenberg1927,Robertson1929} from the translation structure of the one-particle theory without assuming it as an independent axiom. We now ask how that result changes when both position and momentum are interpreted as physical observables of the nonlocal field theory. The distinction is important: the canonical operators $X$ and $P$ determine the underlying quantum kinematics, while the experimentally realized observables may possess finite nonlocal response widths.

The last result concerns the canonical operators $X$ and $P$. Nonlocality does not require their algebra to be deformed as $P$ remains the generator of translations and $[X,P]=i\hbar$. What changes is the physical realization of position and momentum. In NLQFT, point-supported observables are replaced by finite-resolution observables inherited from the nonlocal field algebra. For normalized and centered response kernels, their first moments remain unchanged, while their variances acquire intrinsic nonlocal contributions shown by $(\Delta X_M)^2=(\Delta X)^2+\sigma_X^2$ and $(\Delta P_M)^2=(\Delta P)^2+\sigma_P^2$~\cite{ThompsonLocalization,ThompsonPhaseSpace}. Combining these relations with the canonical Heisenberg inequality gives the physical nonlocal phase-space relation:
\begin{equation}
\left[(\Delta X_M)^2-\sigma_X^2\right]
\left[(\Delta P_M)^2-\sigma_P^2\right]
\geq \frac{\hbar^2}{4}.
\end{equation}

So NLQFT does not replace the Heisenberg uncertainty principle, rather it enlarges the operational uncertainty associated with physical observables. In particular, the position response has an irreducible width:
\begin{equation}
\Delta X_M\geq \sigma_X
=O(L_M)
=O\left(\frac{\hbar c}{E_M}\right),
\end{equation}
thus arbitrarily large canonical momentum spread cannot produce arbitrarily sharp physical localization \cite{ThompsonLocalization}. More generally, when both $\sigma_X$ and $\sigma_P$ are nonzero, exact point-supported phase-space data are replaced by a finite nonlocal phase-space cell \cite{ThompsonPhaseSpace}.

\emph{The Born Rule Without Fundamental Locality} -- The preceding construction suggests that the probabilistic interpretation of the one-particle theory should also be formulated without first introducing fundamentally sharp local observables. In ordinary quantum mechanics the Born rule is commonly written in the position representation as $p(x)=|\psi(x)|^2$~\cite{Born1926}. Taken literally this expression refers to an ideal measurement of position with arbitrarily fine spatial resolution. Such an interpretation is not fundamental in the nonlocal theory developed here. At finite $(E_M)$, exact point-supported localization is not a physically realizable observable, and the fundamental measurement structure must therefore be nonlocal from the outset.

The appropriate starting point is a physical localization observable represented by a positive-operator-valued measure~\cite{DaviesLewis1970,Busch1999}:
\begin{equation}
E_X^{(M)}(dx)=dx\int_{\mathbb R}dy
\kappa_X(x-y)|y\rangle\langle y|,
\end{equation}
where $\kappa_X$ is the normalized nonlocal response kernel induced by the underlying field theory:
\begin{equation}
\kappa_X(x)\geq0,
\qquad
\int_{\mathbb R}dx,\kappa_X(x)=1.
\end{equation}
For a centered kernel we additionally have:
\begin{equation}
\int_{\mathbb R}dx,x\kappa_X(x)=0.
\end{equation}
The states $|x\rangle$ appearing in this representation should not be interpreted as physically preparable states of exact localization~\cite{Busch1999}. They are generalized basis vectors used to represent the canonical one-particle Hilbert space. The physical observable is the complete nonlocal operator $E_X^{(M)}$, not the formal projector $|x\rangle\langle x|dx$.

For a normalized state $|\psi\rangle$, the probability assigned to an outcome contained in the infinitesimal interval d$x$ is therefore~\cite{DaviesLewis1970} $\langle\psi|E_X^{(M)}(dx)|\psi\rangle$. Substituting the nonlocal localization observable gives us:
\begin{equation}
dP_M(x)
=
\langle \psi | E_X^{(M)}(dx) | \psi \rangle
=
dx \int_{\mathbb{R}} dy\,
\kappa_X(x-y)\,|\psi(y)|^2.
\end{equation}
Thus the physically observable probability density is given by $\int_{\mathbb R}dy \;\kappa_X(x-y)|\psi(y)|^2$
or, equivalently can be written as $\kappa_X*|\psi|^2$. The quantity $|\psi(x)|^2$ is therefore not, at finite nonlocality, the probability density associated with a physically realizable measurement at a mathematical point. Rather, it is the canonical coordinate-space density from which the observable probability distribution is reconstructed through the nonlocal response kernel.

This distinction preserves the probabilistic consistency of the theory. Since both $\kappa_X$ and $|\psi|^2$ are nonnegative $p_M(x)\geq0$. Moreover we note that:
\begin{equation}
\begin{aligned}
\int_{\mathbb R}dx,p_M(x)
&=
\int_{\mathbb R}dx
\int_{\mathbb R}dy,
\kappa_X(x-y)|\psi(y)|^2\
\\&=
\int_{\mathbb R}dy,|\psi(y)|^2
\int_{\mathbb R}dx,\kappa_X(x-y)\
\\&=
\int_{\mathbb R}dy,|\psi(y)|^2\
\\&=1,
\end{aligned}
\end{equation}
this tells us that the nonlocal observable preserves positivity and normalization of probabilities.

The same structure holds for momentum~\cite{DaviesLewis1970}. Let the physical momentum observable be:
\begin{equation}
E^{(M)}_{P}(dp)
=dp\int_{\mathbb R}dq,
\kappa_P(p-q)|q\rangle\langle q|,
\end{equation}
where $|q\rangle$ denotes the generalized eigenstate of the canonical
translation generator $P$ with eigenvalue $q$, and $\kappa_P$ is the
normalized physical momentum-response kernel. The observable momentum probability density is then:
\begin{equation}
dP_M(p)
=
\langle\psi|E^{(M)}_{P}(dp)|\psi\rangle
=\int_{\mathbb R}dq,
\kappa_P(p-q)|\widetilde{\psi}(q)|^2.
\end{equation}
Equivalently, the observable momentum probability density is given by:
\begin{equation}
p_M(p)
=
\int_{\mathbb R} dq\,
\kappa_P(p-q)\,
|\widetilde{\psi}(q)|^2
=
\bigl(\kappa_P*|\widetilde{\psi}|^2\bigr)(p).
\end{equation}
Thus neither position nor momentum is fundamentally assigned probabilities at an exact mathematical phase-space point. The state is represented canonically on the Hilbert space, while observable outcome statistics are obtained only after the state is evaluated against the nonlocal physical observables.

The important conclusion we draw is not that the quantum probability law is destroyed by nonlocality. Rather, the object to which that probability law is applied is fundamentally nonlocal. Exact point probabilities arise only when the nonlocal response width is neglected. In this sense the familiar Born rule for position is recovered rather than assumed as a fundamental statement about probabilities at mathematical points.

It is important to distinguish this result from a derivation of the abstract quantum probability functional itself. The assignment $\operatorname{Tr}\left[\rho E_M(\Omega)\right]$ belongs to the Hilbert-space quantum structure inherited from the parent quantum field theory~\cite{DaviesLewis1970}. What the present construction establishes is that, once this quantum structure is reduced to the nonlocal one-particle sector, the physically relevant effects $(E_M(\Omega))$ are nonlocal and finite-resolution. The standard expression $(p(x)=|\psi(x)|^2)$ is then recovered only in the local limit.

The completeness question is therefore sharpened once more:

\emph{A local quantum-mechanical description is incomplete at finite $(E_M)$ not because its probability interpretation fails, but because it identifies the alternatives appearing in that probability interpretation with arbitrarily sharp local observables. The more fundamental theory assigns probabilities directly to nonlocal physical effects, with ordinary pointwise Born probabilities emerging only as their infrared approximation.}

\emph{Quantum Mechanics Without Fundamental Locality} -- We can now return to the question posed in the title. The conclusion is not that quantum mechanics is incomplete as a probabilistic theory, nor that the wave function must be supplemented by hidden classical variables. The underlying relativistic quantum field theory is nonlocal at sufficiently short distances, then an exactly local quantum-mechanical description cannot itself be fundamental. It is instead the infrared one-particle limit of a more general theory in which interactions and physical phase-space observables are nonlocal.

There is also a historical reason why this interpretation is natural within quantum theory. Planck showed that classical assumptions about continuously exchangeable energy could not simply be retained at microscopic scales~\cite{Planck1901}. Einstein treated energy quanta as physically meaningful rather than as bookkeeping devices~\cite{Einstein1905}. De Broglie associated wavelength with matter~\cite{deBroglie1924}, replacing the classical separation between particles and waves by a structure in which both descriptions were necessary. Schr\"odinger made the wave amplitude dynamical~\cite{Schrodinger1926}, while Born identified its modulus squared with probabilities for measurement outcomes~\cite{Born1926}. Heisenberg reorganized mechanics around observable quantities and showed that the simultaneous classical specification of conjugate variables cannot be retained~\cite{Heisenberg1925,Heisenberg1927}. Bohr emphasized that statements about quantum systems acquire physical meaning only in relation to experimentally definable conditions~\cite{Bohr1928}.

Their common lesson is that quantum theory repeatedly forced physics to distinguish mathematical structure inherited from classical mechanics from physical structure that can actually be assigned to a quantum system.

The distinction between the canonical and physical observables is therefore that $X_M=X$ and $P_M=P$ at the level of first moments, while $X_M^{(2)}=X^2+\sigma_X^2 I,\qquad P_M^{(2)}= P^2+\sigma_P^2 I$~\cite{ThompsonLocalization,ThompsonPhaseSpace}. So nonlocality leaves the global symmetry generators intact but changes the operational phase-space resolution of the theory~\cite{Thompson2026Covariance,ThompsonLocalization,ThompsonPhaseSpace}.

The answer to the title is conditional, \emph{local quantum mechanics is complete only as an infrared description}. What becomes incomplete is not quantum mechanics itself, but the additional assumption that its physical observables remain exactly local and arbitrarily sharp in phase space at all scales. Locality becomes an emergent low energy approximation, rather than an axiom imposed at arbitrarily short distances.

\emph{Acknowledgments} -- Research at the Perimeter Institute for Theoretical Physics is supported by the Government of Canada through Industry Canada and by the Province of Ontario through the Ministry of Research and Innovation (MRI).

\end{document}